\documentclass[aps,prb,floatfix,twocolumn,superscriptaddress]{revtex4}
\usepackage{graphicx,amsmath,amssymb,subfigure,epsfig,psfrag,color,verbatim,array}
\providecommand{\mean}[1]{\left\langle #1 \right\rangle}

\newcommand{\myhrulefill}{}

\begin{document}
\title{Noise Predictions for STM in Systems with Local Electron Nematic Order}
\author{Y.~L.~Loh}
\affiliation{Department of Physics, The Ohio State University, Columbus, OH  43210}
\affiliation{Department of Physics, Purdue University, West Lafayette, IN  47907}
\author{E.~W.~Carlson}
\affiliation{Department of Physics, Purdue University, West Lafayette, IN  47907}
\author{K.~A.~Dahmen}
\affiliation{Department of Physics, University of Illinois, Urbana-Champaign, IL 61801}
\date{\today}
\begin{abstract}
We propose that thermal noise in local stripe orientation should be readily detectable
via STM on systems in which local stripe orientations are  strongly affected by quenched disorder.
Stripes, a unidirectional, nanoscale modulation of electronic charge, are strongly affected by
quenched disorder in two-dimensional and quasi-two-dimensional systems. 
While stripe orientations tend to lock to major lattice directions, dopant disorder
locally breaks rotational symmetry.  In a host crystal with otherwise $C_4$ rotational symmetry,
stripe orientations in the presence of quenched disorder map to the random field Ising model.
While the low temperature state of such a system is generally a stripe glass in two dimensional
or strongly layered systems, as the temperature is raised, stripe orientational fluctuations
become more prevalent.  We propose that these thermally excited fluctuations should be
readily detectable in scanning tunneling spectroscopy as {\em telegraph noise} in the high
voltage part of the local $I(V)$ curves.
We predict the spatial, temporal, and thermal evolution of such noise,
including the circumstances under which such noise is most likely to be observed.
In addition, we propose an in-situ test, 
amenable to any local scanning probe,
for assessing whether such noise is due to correlated fluctuations rather than independent switchers.
\end{abstract}
\maketitle
	
\myhrulefill\section{Introduction}

There is experimental evidence that many strongly correlated electronic systems such as 
nickelates, cuprates, and manganites exhibit some degree of local inhomogeneity,
{\em i.e.}, nanoscale variations in the local electronic properties. 
Describing the electronic behavior of these materials involves several degrees of freedom, including orbital, spin, charge, and lattice degrees of freedom. 
Disorder only compounds the problem. Not only can certain types of disorder destroy 
phase transitions, leaving mere crossovers in their wake, it can fundamentally alter ground 
states, sometimes forbidding long range order. Especially in systems where different physical 
tendencies compete, disorder can act as nucleation points for competing ground states. 
The interplay between many degrees of freedom,  strong correlations,
and disorder can lead to  a hierarchy of 
length scales over which the resulting physics must be described. 
While such electronic systems are highly susceptible to pattern formation at the nanoscale, 
{\em unfortunately most of our theoretical and experimental tools are designed for understanding 
and detecting homogeneous phases of matter.}

Because of these difficulties, the presence of quenched disorder in real materials can
make it exceedingly difficult to discern the character of the local pattern formation.
One might expect to overcome the difficulty by growing cleaner samples, but certain
forms of local order (such as stripe orientations, {\em i.e.}, the nematic component of stripes)
are so fragile that any finite disorder precludes a long range ordered ground state
in two dimensions.    (Finite but small disorder can preclude it in layered systems.)
It is therefore desirable to develop theoretical and experimental techniques
aimed at characterizing and detecting local electronic order.  In previous work, we have
shown that noise and nonequilibrium effects can be used to help illuminate the local
electronic tendencies.\cite{rfim-prl}
In this paper, we discuss the implications of such noise for
scanning tunneling microscopy.  In particular, we propose that thermally excited switching
of local stripe orientations should be readily detectable via {\em scanning noise microscopy},
appearing as telegraph noise in the high voltage part of atomic scale $dI/dV$ curves.

Long-range stripe order has been detected in a subset of 
cuprates, where it can coexist with superconductivity,
as well as in the related nickelates.\cite{tranquada-review}
However, most cuprate superconductors lack evidence of long-range stripe
order.  This calls into question theories of cuprate superconductivity
based on quasi-one-dimensional electronic structure.  On the other hand,
because the pairing scale is large, it can in principle be established on short length scales.
(In fact, STM experiments have shown that the correlation length of
the superconducting order parameter, {\em i.e.}, the distance over which the
gap scale can vary, is about 10-20\AA.)  Thus the salient question is
not whether long-range stripe order is ubiquitous in cuprate superconductors,
but whether short-range stripe order is ubiquitous.  
Nevertheless,  the fact that long-range ordered stripes have been observed in several
materials with strong electronic correlations means that the study of 
these structures in the presence of quenched disorder is interesting in its own right,
regardless of issues of the mechanism of cuprate superconductivity.

Indeed, recent STM experiments show evidence of an
electronic cluster glass of locally unidirectional domains
in Na-CCOC and Dy-Bi2212 at low temperature.\cite{seamus-glass}
We predict that upon raising the temperature in such 
systems, more and more stripe domains will become thermally
excited, resulting in local fluctuations in the stripe orientation.  
The main purpose of this paper is to predict the spatial, temporal, and 
thermal evolution of such noise.

Telegraph noise in the mesoscopic
transport properties of a YBCO nanowire has been reported by
Bonetti {\em et al.} in the pseudogap regime\cite{bonetti-04}, appearing on timescales
of about 10-50 seconds, and producing resistance fluctuations of the 
order of $0.5\%$ of the total resistance of the wire.
We have previously modeled this telegraph noise via a mapping of stripes in the
presence of quenched disorder to the random field Ising model\cite{rfim-prl}
(as explained below), along with a further mapping to a related
random resistor network.  While the nanowire transport results can be
captured by this model, it was not possible to definitively rule out other sources
of the telegraph noise, such as superconducting fluctuations.\cite{bonetti-04}
Transport noise has also been reported on timescales of 1-1000 seconds in the $c$-axis resistivity of bulk LSCO at $T \lesssim 0.3$K by Raicevic and coworkers.\cite{raicevic-2008} 

Is the low temperature glass of unidirectional domains observed in
Na-CCOC and Dy-Bi2212 related to the 
transport noise observed in YBCO and LSCO?	
If similar noise were to be found with STM, it would begin to bridge the gap
between the real-space imaging capability of that probe and the noise
seen in mesoscopic and macroscopic systems.  This would allow
concrete connections to be made between the microscopic and macroscopic
behavior, and provide guidance as to how to extract information about local
order from macroscopic probes when microscopic probes such as
STM are not feasible.    Furthermore, since transport noise is a bulk measurement,
finding evidence of similar noise in STM would
establish a connection between physics happening in the bulk
and that occurring on the surface.
In addition, the wealth of information available from STM can be used to set
the parameters of our  model, so that concrete predictions can then
be made for other measurements.

\begin{figure}[thb]
\centering
\subfigure[~Two Dimensions]{
	\includegraphics[width=0.45\columnwidth]{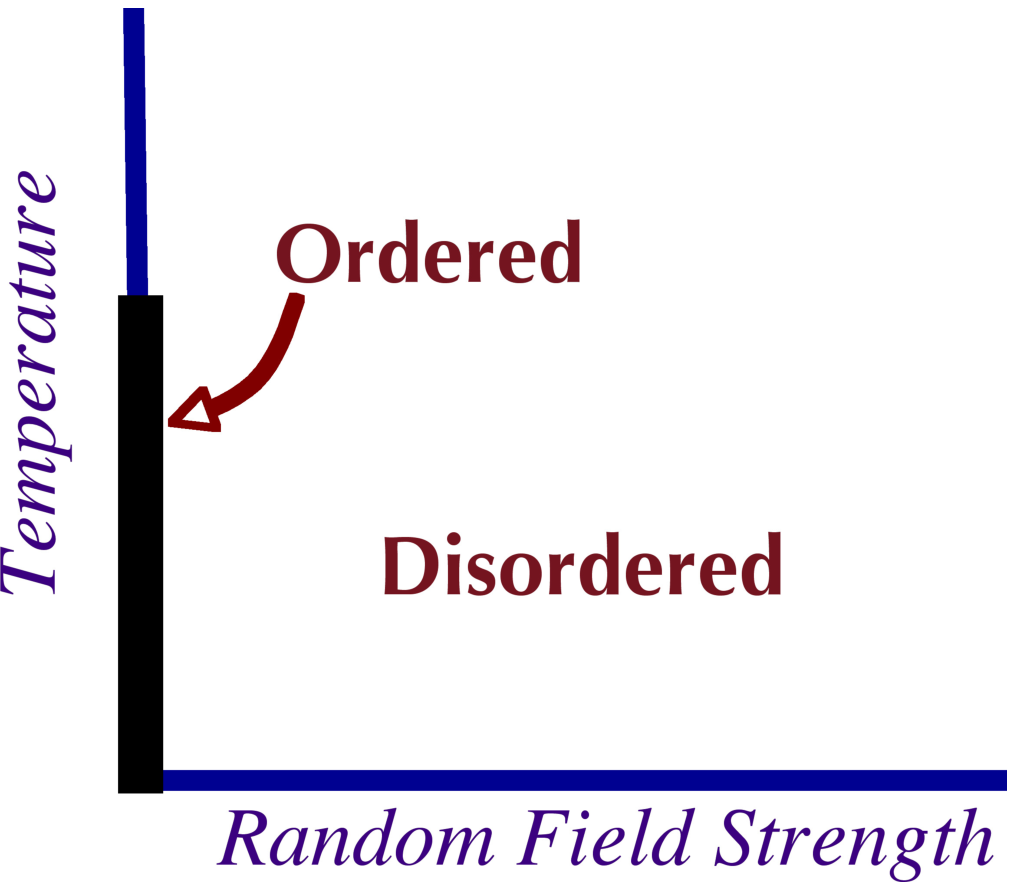}
	\label{f:2D-RFIM}
}
\hspace{0.01\columnwidth}
\subfigure[Three Dimensions]{
	\includegraphics[width=0.45\columnwidth]{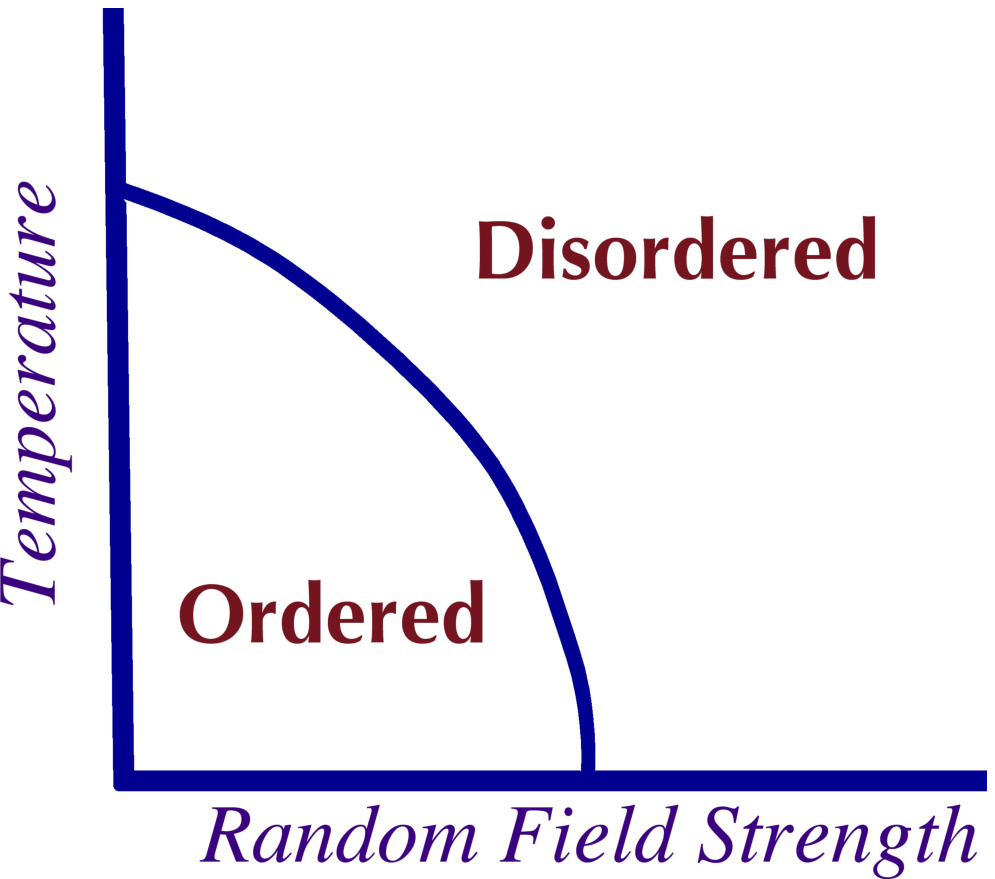}
	\label{f:3D-RFIM}
}
\caption{
Schematic phase diagram of the random field Ising model.  (a) In two dimensions, the
critical disorder strength is zero, and for any finite disorder strength the model is in a disordered phase
at all temperatures.  (b)  In three dimensions, there is a finite critical disorder strength.
\label{f:phasediagram}
}
\end{figure}
\begin{figure}[!htb]
\centering
\subfigure[Color visualization]{
	\includegraphics[height=0.35\columnwidth]{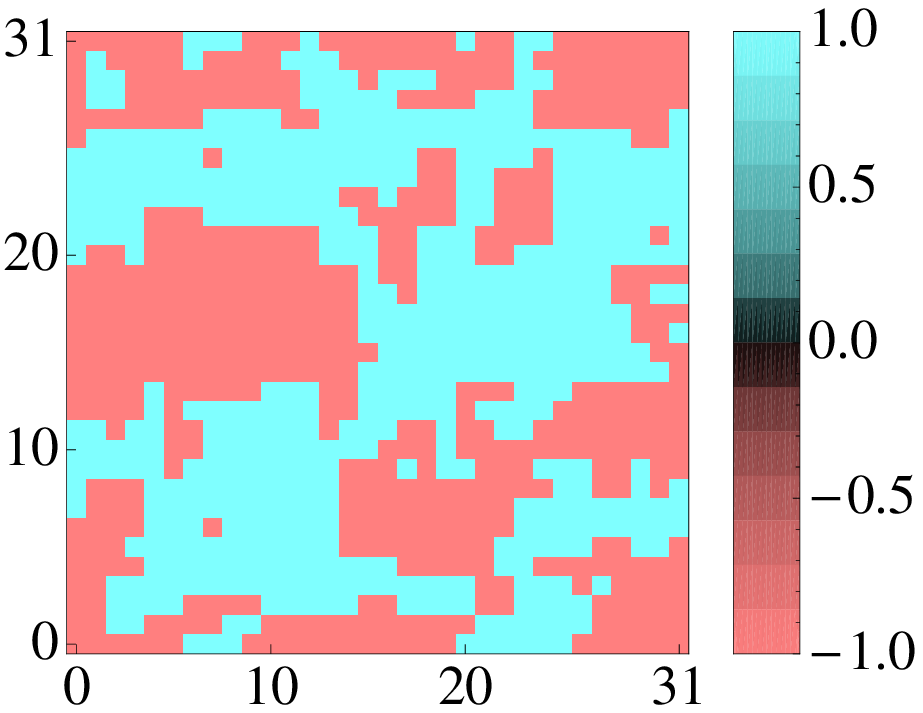}
	\label{f:snapshot}
}
\hspace{0.01\columnwidth}
\subfigure[Stripe visualization]{
	\includegraphics[height=0.35\columnwidth]{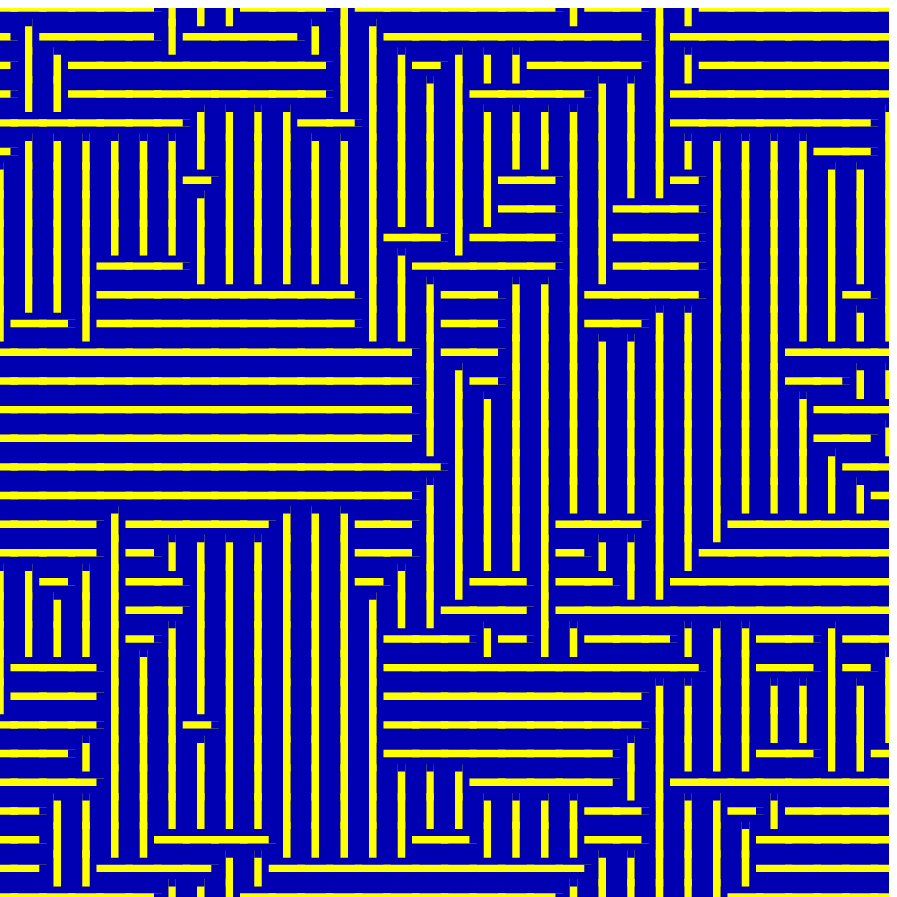}
	\label{f:snapshot-stripes}
}
\caption{
(Color online) A snapshot of a layered random-field Ising model with $R=2.5J$, $T=0.5J$ on a $32 \times 32 \times 8$ cubic lattice (see text for details).  Only the top layer is shown.  (a) Here, the two states of the Ising variables, $\sigma_i=\pm 1$, are represented by red and blue patches.  (b) The same results visualized as vertical and horizontal stripes.
\label{f:snapshots}
}
\end{figure}

We now describe the mapping of stripe orientations in the presence of quenched disorder
to the random field Ising model.\cite{rfim-prl}
Stripes in a host crystal such as the cuprates and nickelates tend to lock to favorable lattice directions, often either vertically or horizontally along Cu-O (Ni-O) bond directions.  
This breaks the symmetry of the host crystal from 4-fold to 2-fold rotational symmetry.  With two allowed orientations, the nematic component of the stripe order parameter is of Ising character, where the Ising variable or ``pseudospin''
$\sigma = \pm 1$ corresponds to horizontal or vertical stripe patches.  The tendency of neighboring stripe patches to align is modeled as a ferromagnetic interaction between nearest-neighbor Ising variables.  Dopant atoms act like a random field on the electronic stripe nematic, so that in any given region, the particular dopant arrangement will favor one or the other orientation of the stripes locally, producing a disorder-dependent pattern of stripe patches with two orientations.  Thus stripes in a host crystal with dopant disorder can be mapped to the random-field Ising model (RFIM):\cite{rfim-prl}
\begin{equation}
H = -J \sum_{\left<ij\right>_{||}} \sigma_i \sigma_j - J_{\perp} \sum_{\left<ij\right>_{\perp}} \sigma_i \sigma_j 
-\sum_i (h+h_i) \sigma_i
	\label{e:hamiltonian}
\end{equation}
where $J$ is the in-plane coupling, $J_\perp$ is the interlayer coupling, 
$h$ is an orienting field, and $h_i$ are the random pseudofields at each site.  
In the summations, $\left<ij\right>_{||}$ denotes a summation over sites within a plane,
and $\left<ij\right>_{\perp}$ denotes summation over sites in neighboring planes.
The orienting field may be, for example, $h \propto B_x{}^2 - B_y{}^2$ where $\vec{B}$ is the physical external magnetic field.\footnote{Other examples of an external perturbation which breaks the rotational
symmetry of the lattice and would therefore tend to align stripes include strain, high currents, and uniaxial pressure.}  
The random fields $h_i$ are chosen independently from a Gaussian distribution with standard deviation $R$ (the disorder strength).  The order parameter $M=\frac{1}{N} \sum_i \mean{\sigma_i}$ represents the degree of 
stripe orientation, {\em i.e.}, nematic order.
The size of a stripe patch sets the spacing of the Ising lattice, and we may use information from
experiments to set bounds on this lattice spacing.  For example, for the purposes of comparing with 
STM experiments on Na-CCOC and Dy-Bi2212\cite{seamus-glass}, a reasonable lattice spacing
would be 16\AA.  

The materials of interest are layered materials, corresponding to a 
quasi-two-dimensional system.  
As shown schematically in Fig.~\ref{f:phasediagram}, for any finite disorder strength, the two-dimensional RFIM remains disordered at all temperatures in the thermodynamic limit.\cite{bray-moore-85}
In three dimensions, the RFIM can order below a critical disorder strength.  
For the finite layered systems we consider here, there is a finite critical disorder
strength, although it may be quite suppressed for strongly layered systems.  
This implies that materials which are strongly layered are less likely to
support an ordered Ising nematic phase than materials 
with stronger coupling in the $c$-direction.
Although the charge component of stripes is subject to Coulomb
coupling from plane to plane, because they are overall charge neutral
this coupling is greatly diminished for distances longer than the
local modulation wavelenth (about $16$\AA), and 
the short-range interactions of our model are sufficient to 
describe the $c$-axis coupling.

\begin{figure*}[!htb]
\centering
\subtable{
\begin{tabular}{ 
	m{.08 \textwidth}
	m{.22 \textwidth}
	m{.19 \textwidth}
	m{.22 \textwidth}
	m{.22 \textwidth}   
}
$T=2.0J$ &
\includegraphics [height=35mm] {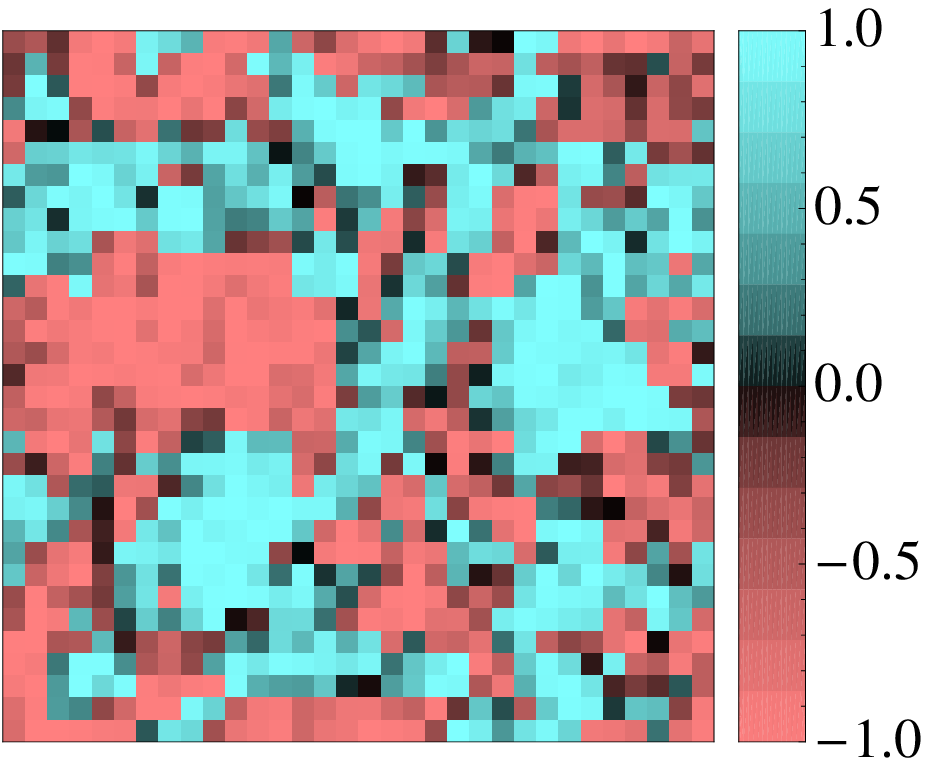}  &
\includegraphics [height=35mm] {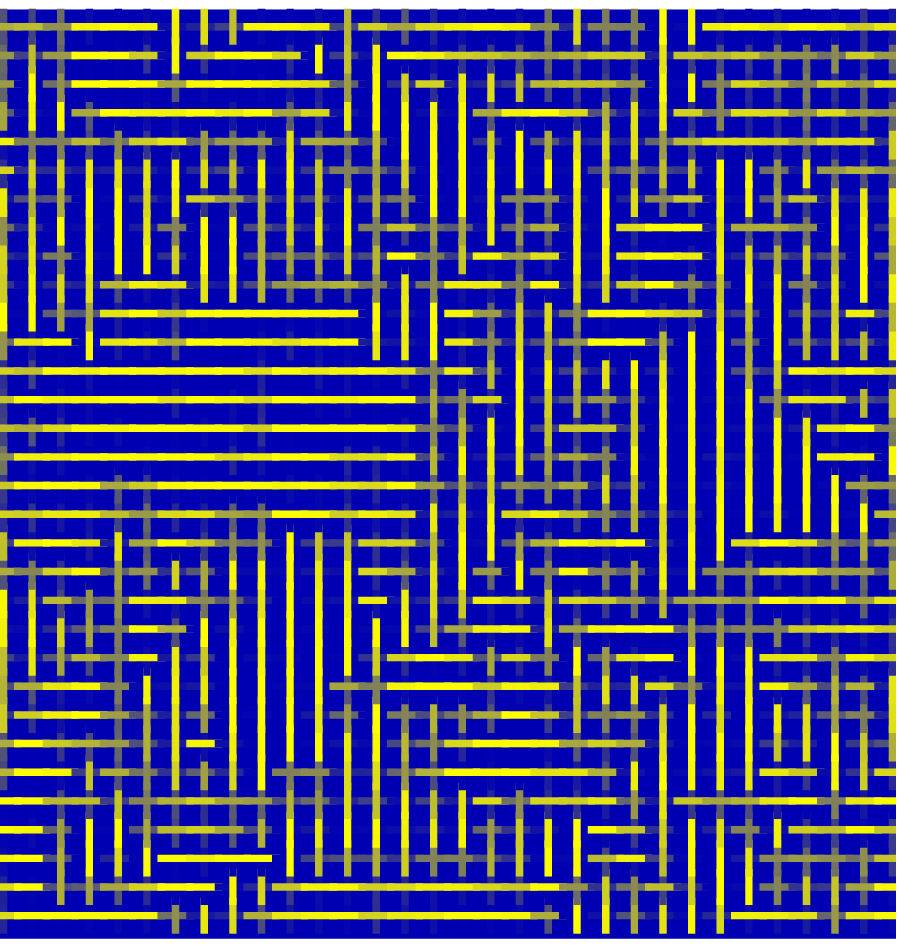}  &
\includegraphics [height=35mm] {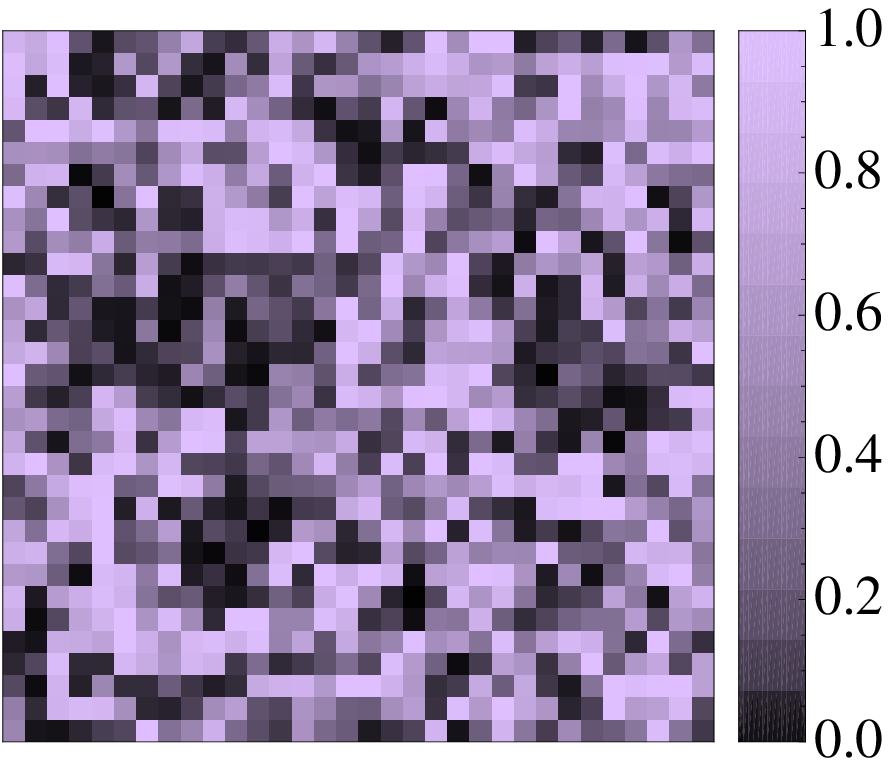} &
\includegraphics [height=35mm] {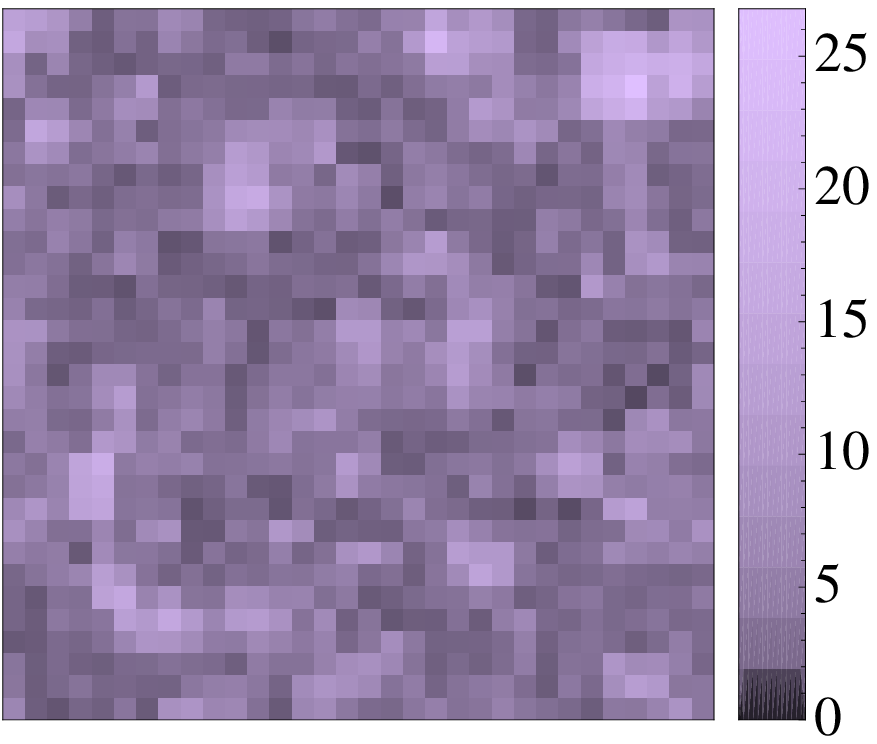}
\\
$T=1.5J$ &
\includegraphics [height=35mm] {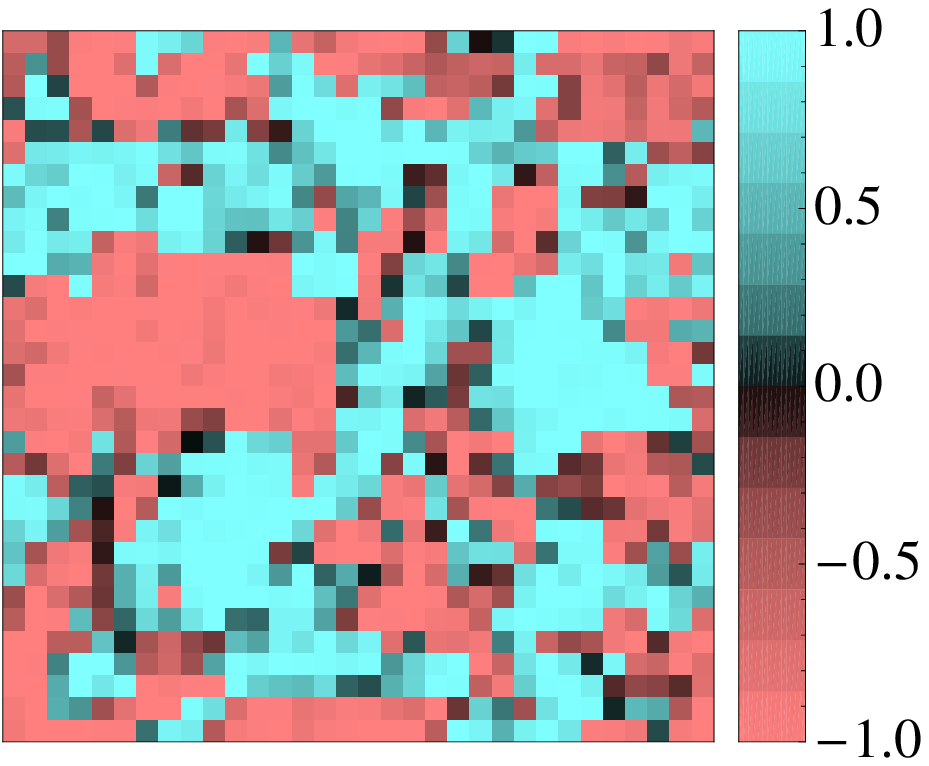}  &
\includegraphics [height=35mm] {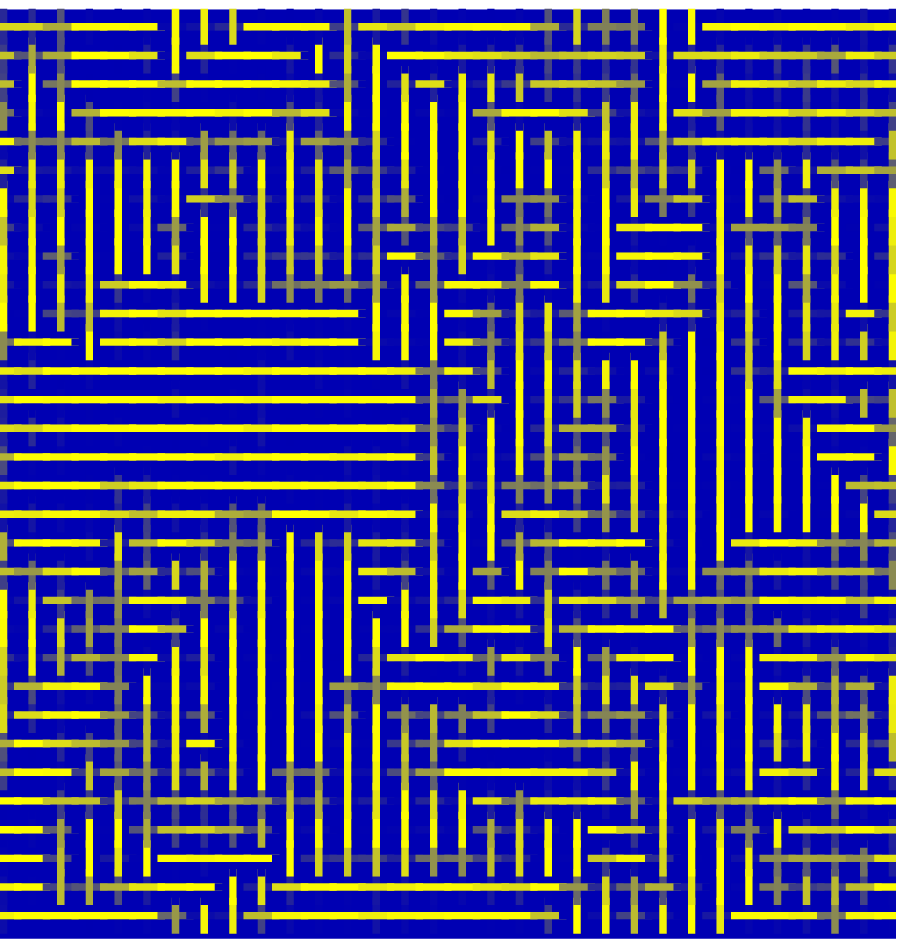}  &
\includegraphics [height=35mm] {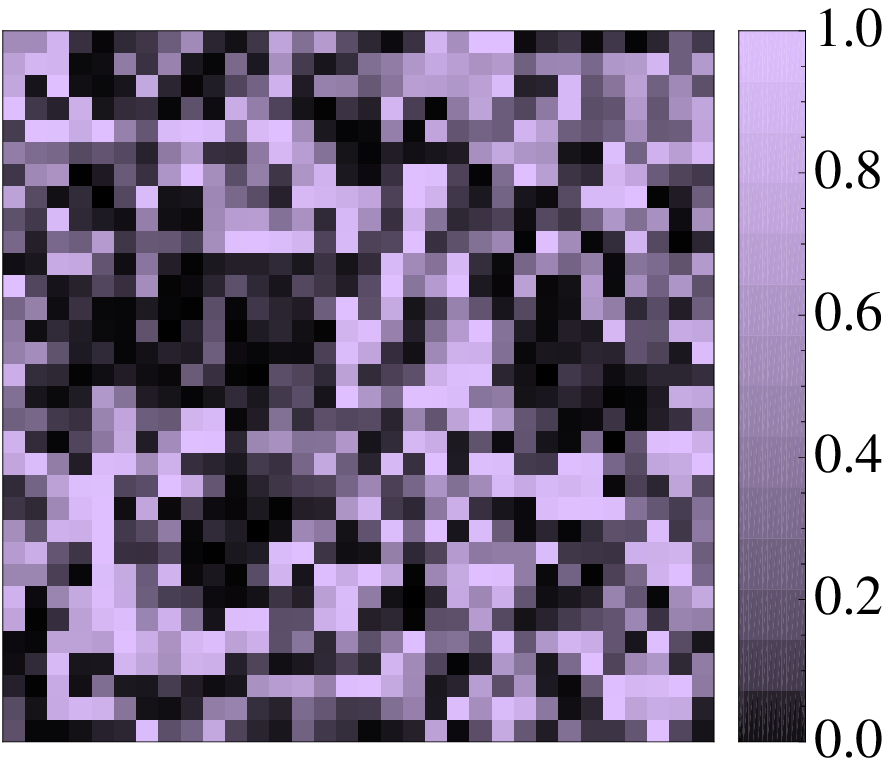} &
\includegraphics [height=35mm] {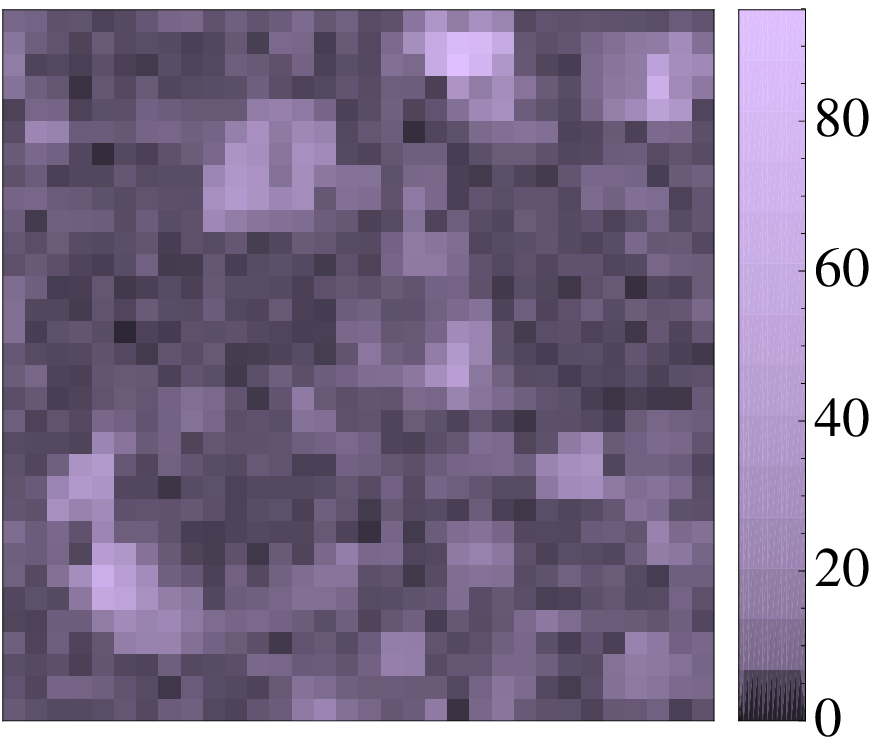}
\\
$T=1.0J$ &
\includegraphics [height=35mm] {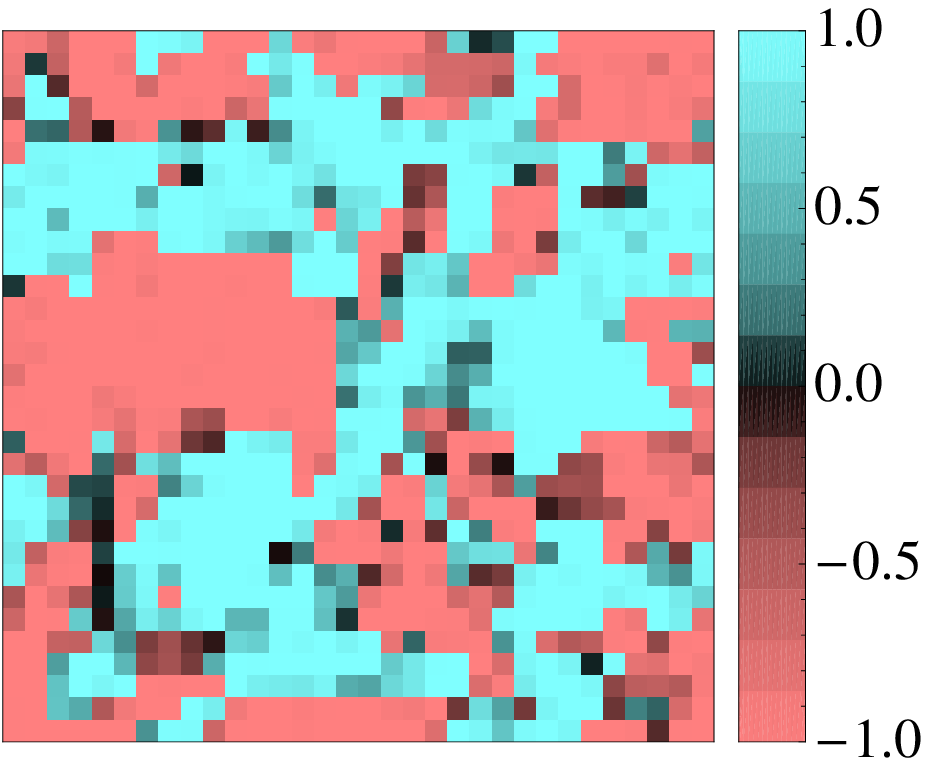}  &
\includegraphics [height=35mm] {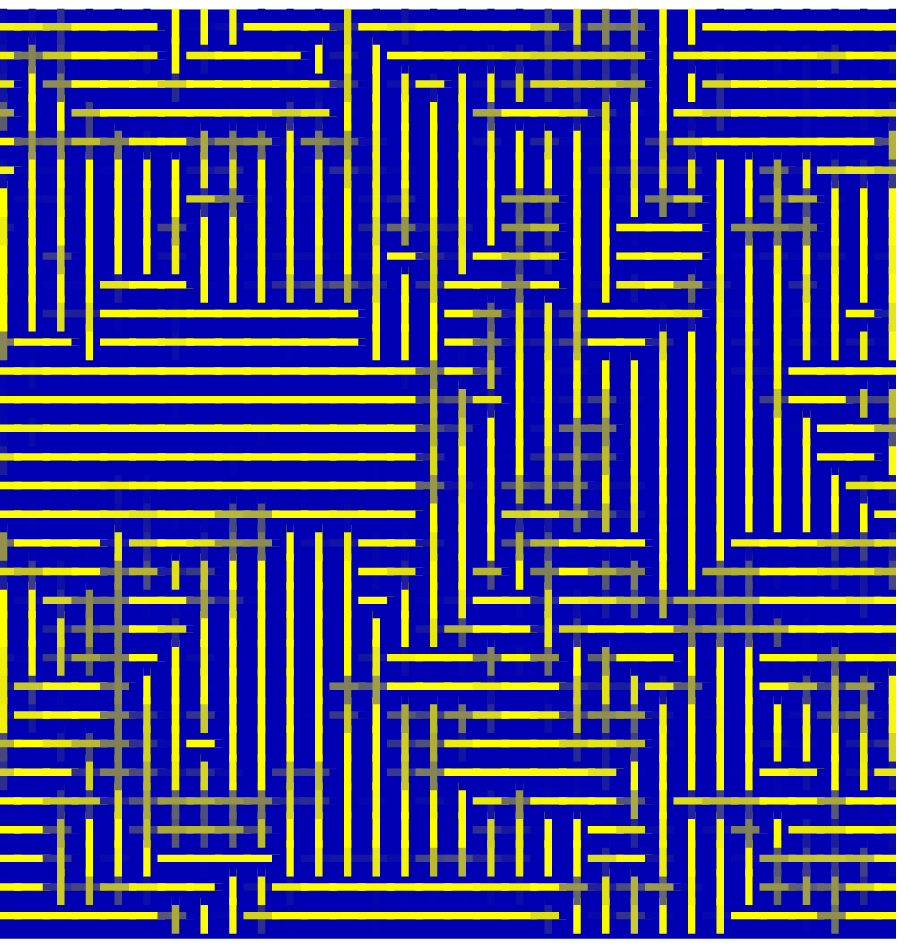}  &
\includegraphics [height=35mm] {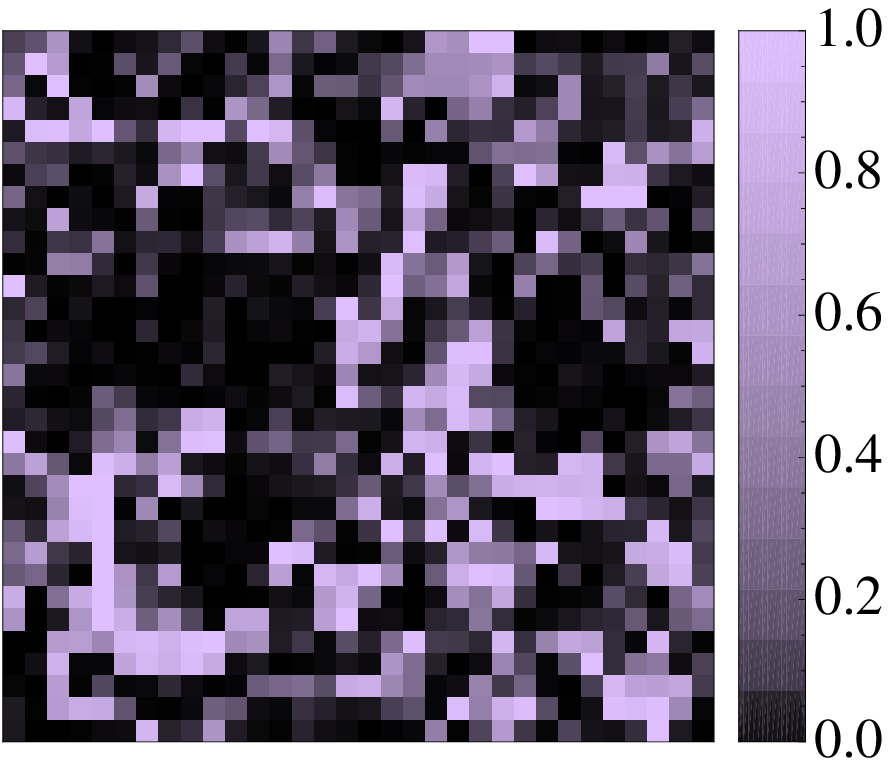} &
\includegraphics [height=35mm] {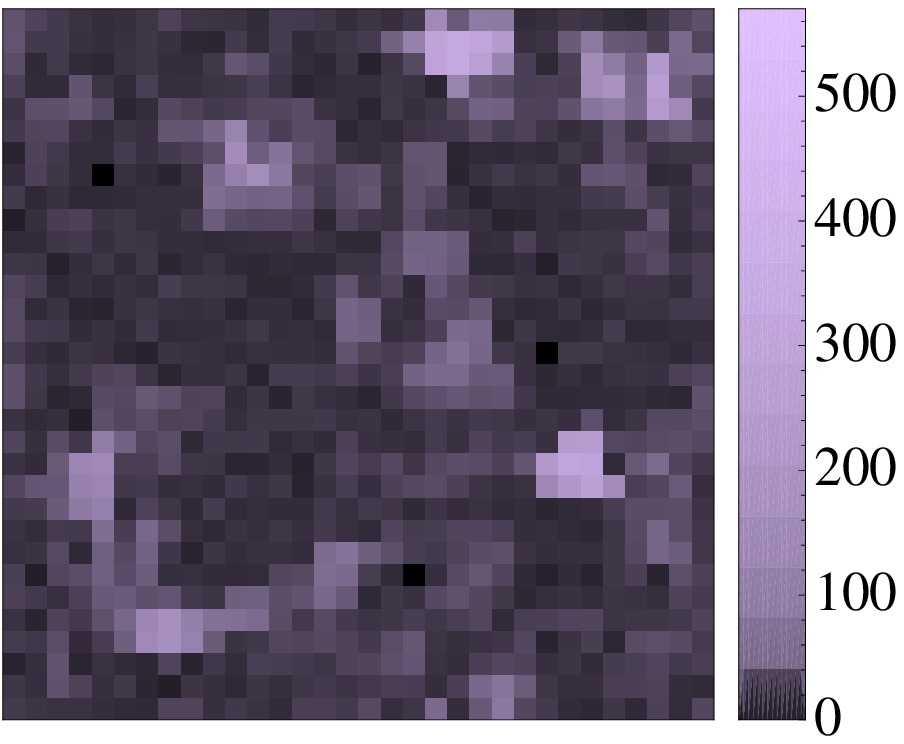}
\\
$T=0.5J$ &
\includegraphics [height=35mm] {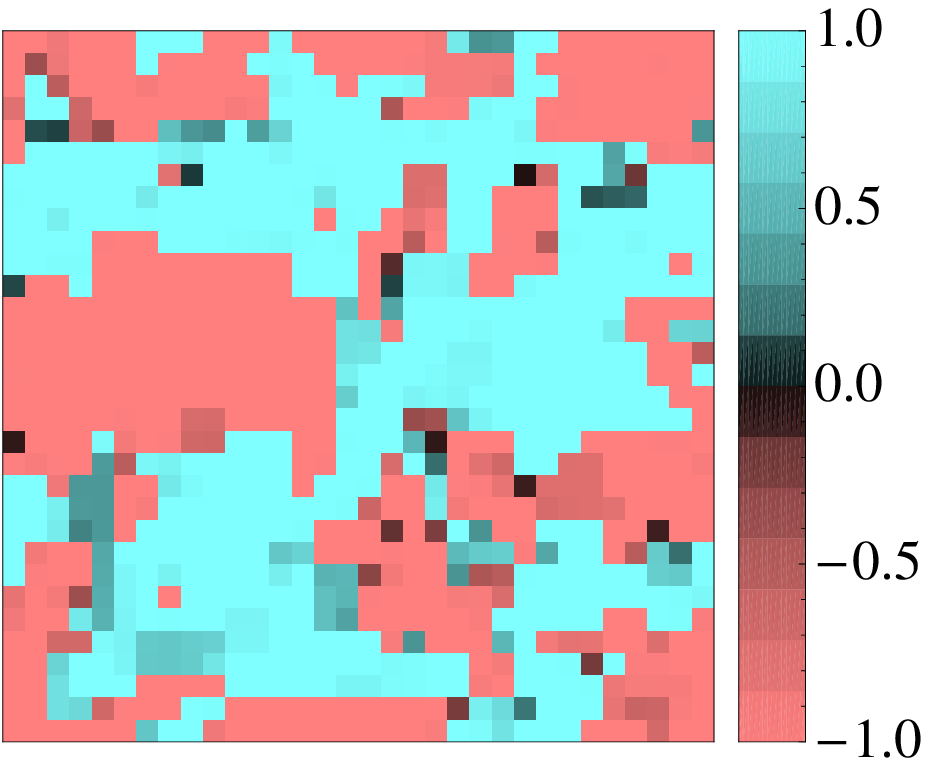}  &
\includegraphics [height=35mm] {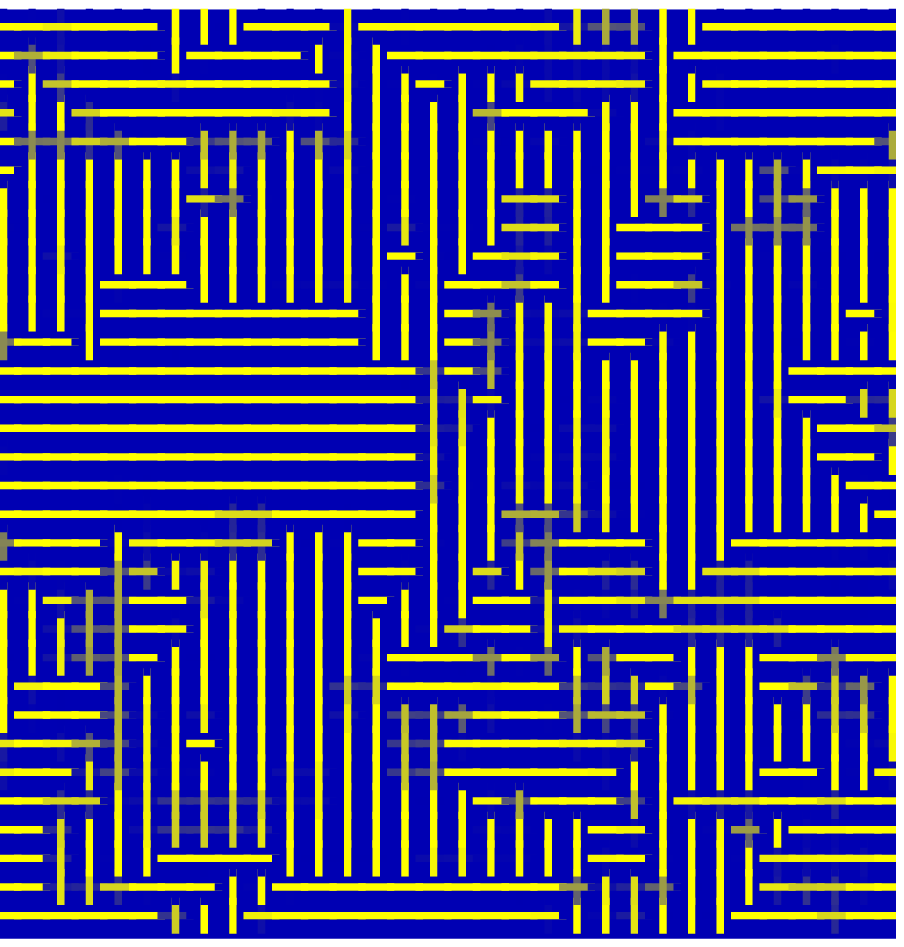}  &
\includegraphics [height=35mm] {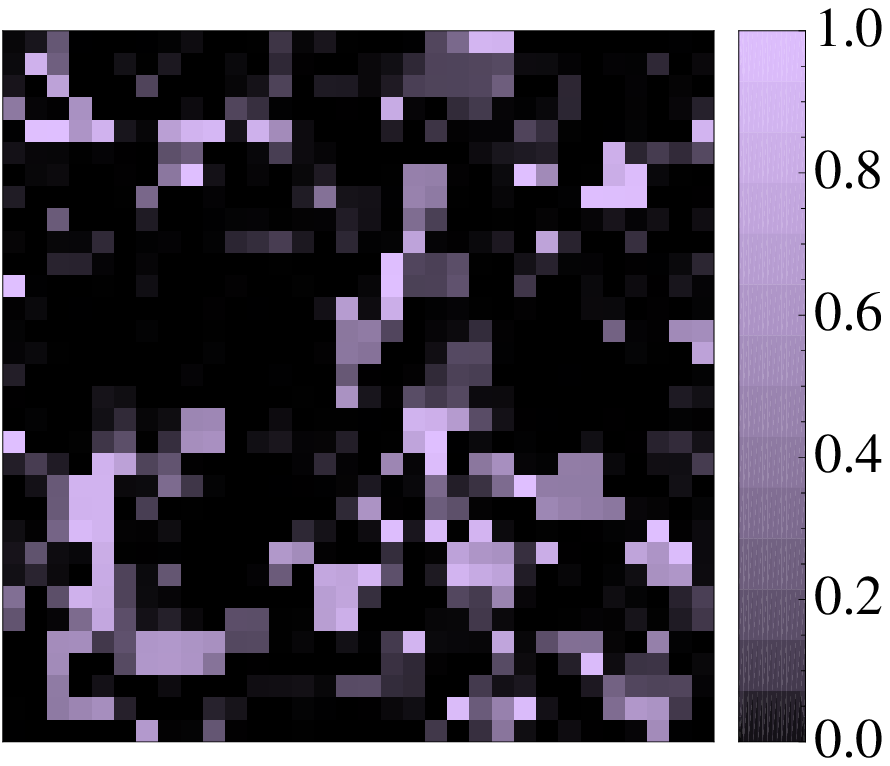} &
\includegraphics [height=35mm] {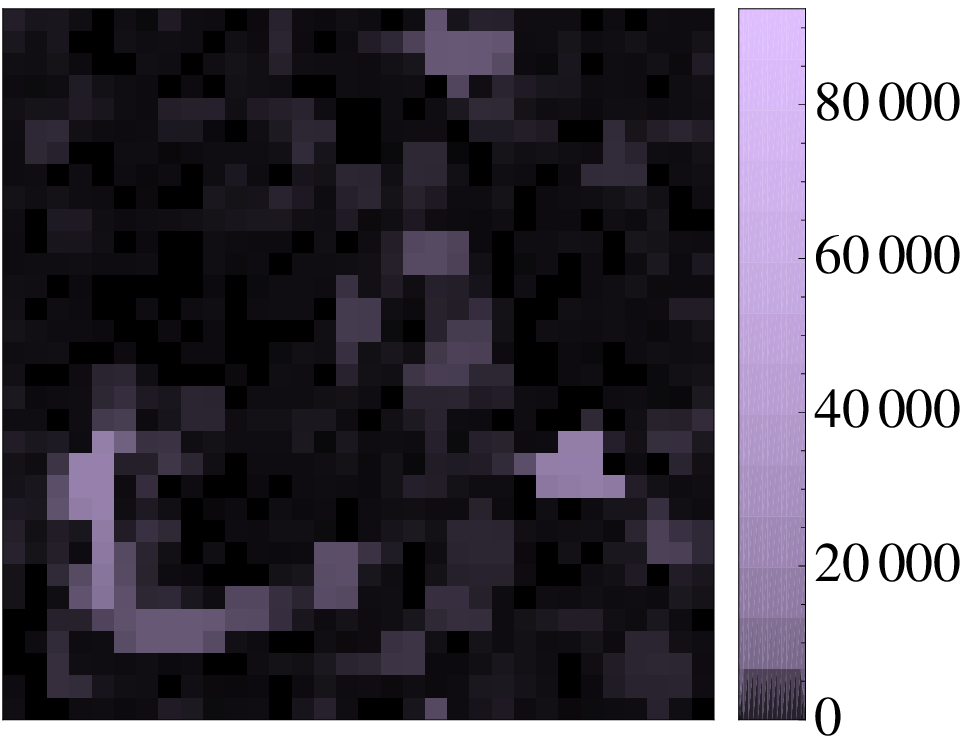}
\end{tabular}
}
\caption{
(Color online) Fluctuations of the RFIM for a disorder strength of  $R=2.5J$ at various temperatures as noted in the figure,
on a $32 \times 32 \times 8$ cubic lattice (see text for details).  Only the top layer is shown.
In the leftmost column, we show the time-averaged (pseudo)spins $\mean{\sigma_i}$, where 
blue and red denote positive and negative averages, respectively, with 
shades of grey denoting average pseudospins near zero.  
In the next column, we plot each averaged pseudospin variable as the director of the Ising nematic,
to visualize the pattern of horizontal and vertical stripes.  
Red sites from the first column correspond
to horizontal stripe orientation, and blue sites from the first column correspond to vertical stripe orientation.
The third column shows the variance of the (pseudo)spins, $\left<(\Delta\sigma_i)^2\right>$.  
The rightmost column shows the local switching timescale $\tau_i$ (see text for definition).  
At low temperatures, fluctuations are all but frozen out, and local switching timescales
diverge.  As the temperature is raised, more and more clusters become active, showing 
as bright spots in the variance maps (third column).  Note that the boundaries of the low temperature
clusters become active first as temperature is raised.  
At high temperature, noise is apparent throughout the sample.
\label{f:kitchen-sink}
}
\end{figure*}

\myhrulefill\section{Thermally activated stripe orientation switching}\label{s:numerics}

In Figure~\ref{f:snapshots}, we show an example of the low temperature
behavior in the disordered phase, revealing a low temperature
pattern with locally unidirectional domains of various sizes.  
Fig.~\ref{f:snapshots}(a) shows a single thermalized configuration in our model,
and Fig.~\ref{f:snapshots}(b) shows the corresponding stripe visualization.
In order to do the calculation, 
we have performed Monte Carlo simulations of Eqn.~\ref{e:hamiltonian} with checkerboard Glauber updates
\footnote{In some magnetic systems Kawasaki dynamics may be more appropriate, as they conserve local magnetization.  However, in our problem there is no reason for nematicity to be locally conserved, so Glauber dynamics are appropriate.
Furthermore, because the switching of the local nematic orientation (a single Ising variable in our coarse-grained model) 
involves many electrons, we do not expect this to be a coherent quantum process.
Therefore, we treat the dynamics classically, rather than using a Caldeira-Leggett model.}
for a layered RFIM on a $32 \times 32 \times 8$ cubic lattice.  We use periodic boundary conditions in the $x$- and $y$-directions along with open boundary conditions in the $z$-direction, and we show results from the
top layer, as a rough approximation to the geometry of an STM experiment on a surface.  
We have used coupling constants appropriate to a layered system, 
$J=1$ within each layer and $J_z=0.125$ between layers,
with a disorder strength $R=2.5J$, which is above the critical disorder strength.
This typical configuration is taken after approximately $10^7$ timesteps.
(It takes about $5 \times 10^4$ steps to thermalize at these parameters.)
In Figure~\ref{f:snapshot}, we show a conventional visualization of 
the Ising state, using color to represent the state of the spin.
In Figure~\ref{f:snapshot-stripes}, we have shown the
director of the Ising nematic, in order to represent the average direction
in which stripes run at any given site in the corresponding Ising model. 
\begin{figure}[thb]
\centering
	\includegraphics[width=0.95\columnwidth]{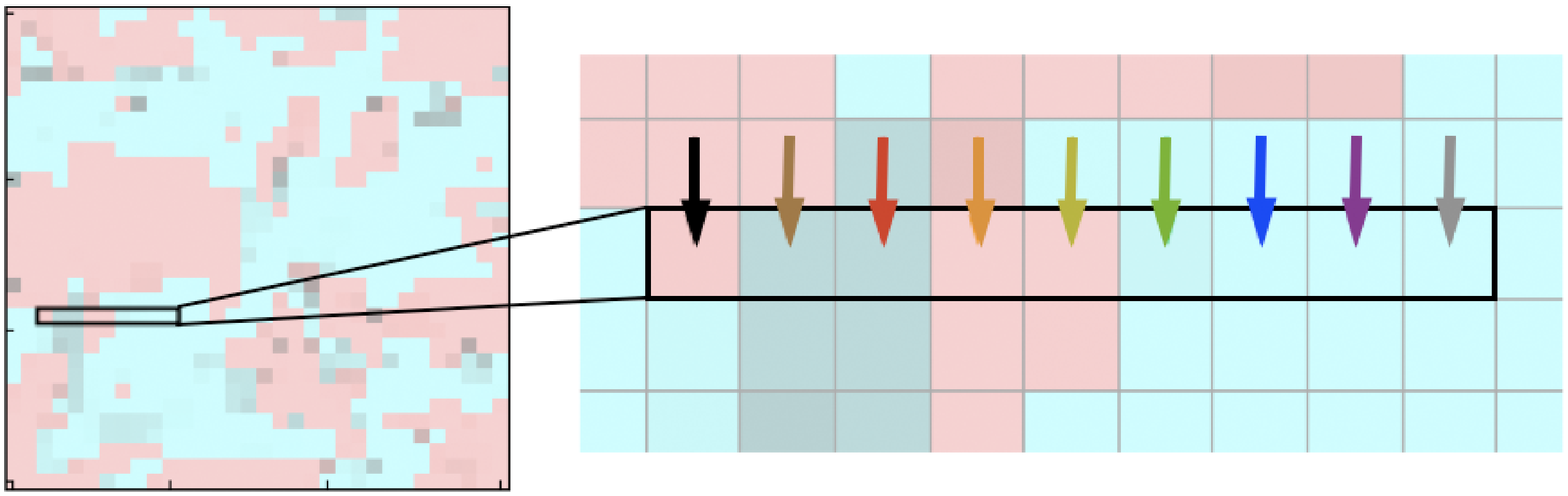}
\caption{
(Color online) 
A line of sites  
in the RFIM at $R=2.5J$ and $T=0.5J$.
This corresponds to the 
simulation from the fourth row of Fig.~\ref{f:kitchen-sink},
and we have chosen a line of sites going through an ``active''
region.  
The colored arrows in the inset correspond to the colors of 
the time traces shown in Fig.~\ref{f:st} and of the block variances
shown in Fig.~\ref{f:blockvar}.
\label{f:index2}
}
\end{figure}

Because we have used a disorder strength which is larger than the critical
disorder strength, the ground state configuration is dominated by quenched
disorder, and although there are domains of various sizes, the average domain size is small. 
A consequence of this is that upon thermal cycling, the
low temperature state returns to the same pattern of stripe orientations,
since the local orientation of each stripe patch is dominated by the
quenched disorder rather than by the thermal history.  
The tunneling 
asymmetry maps of Ref.~\onlinecite{seamus-glass} bear a strong resemblance to configurations of the RFIM (see Fig.~\ref{f:snapshot-stripes}).  
Note that in Fig.~\ref{f:snapshots} we have shown a single (typical) configuration,
rather than an ensemble average.  We believe that this is the correct comparison for 
STM experiments at $T=4.2$K
(as opposed to an ensemble average),
since the timescales over which a single $dI/dV$ curve
is recorded are small compared to the switching time of stripes at that 
temperature.\footnote{Recall that at 100K, telegraph noise in the YBCO nanowire has a timescale of
10-50 seconds.  By 4.2K, thermal fluctuations in stripe orientation should be frozen out in that material.
Likewise, fluctuations in stripe orientation should be similarly frozen out in BSCCO and NCCOC at 4.2K}

\begin{figure*}[thb]
\centering{\includegraphics[width=0.95\textwidth]{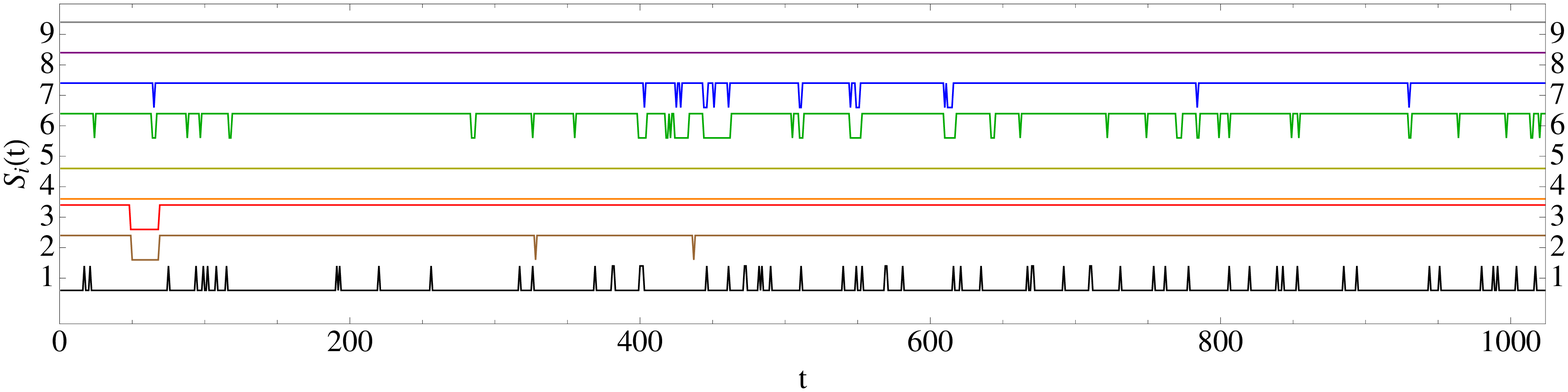}}
\caption{
(Color online) 
Time traces of the nine sites denoted in Fig.~\ref{f:index2},
with $R=2.5J$ and $T=0.5J$.
The colors of the time traces correspond to the colors of the arrows in Fig.~\ref{f:index2}.
\label{f:st}
}
\end{figure*}

While most fluctuations are frozen out at low temperature, upon raising 
the temperature, certain regions of the system become active, and
exhibit thermally driven stripe orientation fluctuations.
In an STM experiment, this should lead to local telegraph noise
in the active regions, where some atomic positions will alternate
between having a charge stripe on them or not.  
We  show below where to look for such noise as a function of
position and temperature, and then describe an algorithm
for detecting whether such noise is due to correlation 
effects rather than from some local, noninteracting source.

Note that microscopically, a rigid rotation is not required in order to produce the thermally fluctuating local Ising variable of our model.  Rather, minor local rearrangements are sufficient to change the direction of any particular unidirectional domain. Exactly how such a local rearrangement transpires is a question of the local energy barriers of a disordered, many-body system, and will be dominated by whatever is the lowest local activation energy in any given region.  Our proposal is independent of these microscopic details. 
It is, of course, conceivable that other mechanisms (e.g., side-to-side stripe fluctuations in a disordered system
or fluctuations of other inhomogeneous patterns) may give rise to fluctuating Ising variables.

Because the materials in question are extremely complicated
and each Ising variable in our model encompasses many electrons,
it is difficult to arrive at a reliable microscopic estimate of timescales.
Nevertheless, here is a rough argument to show that the numbers 
are not unreasonable.  
Assuming that the stripe fluctuations
involve electron dynamics, the attempt frequency would be of the order
of the Fermi velocity (2 eV\AA) divided by the stripe spacing (16
\AA), that is, $10^{15}$Hz.  Noise in the YBCO nanowire was observed on a
timescale of about 50 seconds at 100K.\cite{bonetti-04}  For an Arrhenius type
law, this corresponds to a local energy barrier of about 3800K, not
far from the bare Heisenberg spin exchange couplings in the
cuprates.  While the Heisenberg coupling is not equal to our Ising coupling,
it is one parameter which sets the Ising coupling, and so the two may be of the same magnitude.
Note, however, that due to the randomness in our model, every cluster has a different timescale, 
many clusters have multiple timescales,  and these {\em timescales span several orders of magnitude.} 
As a result,  the issue is less one of whether timescales will be accessible
experimentally, and more one of where spatially to look for fluctuating domains.  
We address this issue in Fig.~\ref{f:kitchen-sink}.

Figure~\ref{f:kitchen-sink} shows what happens when the temperature is raised on the
same system as that in Figure~\ref{f:snapshots}.  
In the leftmost column of Fig.~\ref{f:kitchen-sink}, the grey and black regions are exhibiting stripe orientation fluctuations.
In the corresponding stripe visualization (second column of Fig.~\ref{f:kitchen-sink}), this appears
as places where the different stripe orientations appear to ``invade'' each other.  
These active regions have relatively high variance compared to
the surrounding regions, as shown in the third column of  Fig.~\ref{f:kitchen-sink}.  
The black patches in the leftmost column can be seen as active (white) patches in the variance maps.
As temperature is raised, more and more regions in the sample become ``active''.
The places which become active first are boundaries between the low temperature (frozen) domains,
and especially domain boundaries with several small domains.  
These are the best places to look for local telegraph noise
upon raising temperature.

Of course, the observation of telegraph noise is not sufficient to determine whether 
correlations are present in the system.  Telegraph noise could arise from other sources,
such as an unstable defect on the tip, or even defect hopping on the surface of the
material itself.
However, these are uncorrelated sources of noise due to isolated,
independent switchers.  We are interested in how to identify noise arising from a 
correlated system.  
One way to distinguish such ``spurious'' possible sources of telegraph
noise from that due to correlations in an interacting model such as the one considered here
is to analyze the power spectrum of a time series of the data.  It is well known that 
for a ``stationary'' random telegraph signal ({\em i.e.}, one with
a constant probability of flipping from ``up'' to ``down'', and with a separate unchanging
probability of flipping from ``down'' to ``up'') the resulting power spectrum is a Lorentzian,\cite{weissman-rmp,machlup}
and deviations from this form would presumably indicate correlations.\footnote{Second spectra,
another frequency-domain criterion, may also be used.\cite{weissman-rmp}}
However in the materials of interest, where the timescales have been observed to be as long as
10-50 seconds in YBCO\cite{bonetti-04} and 1000 seconds in LBCO\cite{raicevic-2008}, a time domain criterion is much more desirable.

We first describe what we mean by correlated switchers, and then
we present a criterion for detecting 
spatial correlations using a time domain analysis at a single site.
In Figure~\ref{f:index2}, we show a few sites in a larger system, 
with color-coded arrows denoting particular local stripe patches.
We have chosen some sites which are mostly ``vertical'' (blue), 
and some which are mostly ``horizontal'' (red), along with some sites
which are active switchers (grey).  
The colors of the arrows denoting particular sites in Fig.~\ref{f:index2}
correspond to the colors of the time traces of $\sigma_i$ in Fig.~\ref{f:st}.  
The local orientation at each site displays telegraph noise, but the nature of the telegraph noise is different for different sites.  For example, the red trace shows a few sharp spikes, whereas the blue and green traces show frequent switching.  Notice that the noise is {\em correlated}: when the blue spin is  ``up'', the green spin is also up;  however, when the blue spin is down, the green spin becomes ``active'' and switches rapidly.  This behavior indicates correlated switchers, which is the essence of the interacting model we use.  
One might imagine that such correlations could be detected via simultaneous measurements with two (or more)
STM tips;  however, that would require placing the tips closer together than can be achieved with
current setups.
We therefore focus on 
developing a criterion for detecting correlated noise with a single tip.

One way to characterize the temporal behavior of the local noise is to study the autocorrelation function, $\Psi_i(t) \sim \mean{\sigma_i(0) \sigma_i(t)}$.  
However, we have found that the autocorrelation is prone to large statistical errors at long times, and hence provides an unreliable characterization of the data.  
The origin of this issue lies in the interactions of the system.  Because each site is
interacting with several other sites, each site responds at several different timescales.
The true asymptotic behavior of the local autocorrelation is not achieved until well beyond all
of these timescales.

\begin{figure*}[!htb]
\centering
\includegraphics[width=0.98\textwidth]{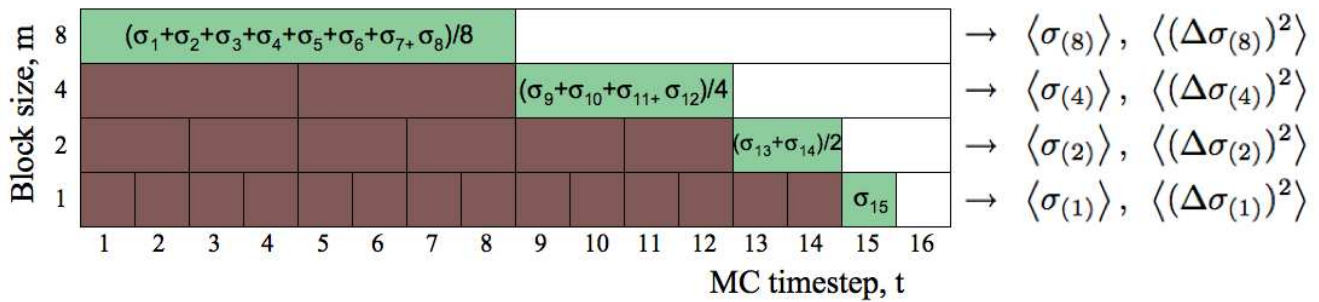}
\caption{
(Color online) Visualization of the ``real-time'' reblocking algorithm for the spin history at a single site.  The horizontal axis represents Monte Carlo time (or real time in a discrete time series from experiment); 
the vertical axis represents the block size, $m$. 
Dark areas represent previously computed blockspin values that have already have been discarded from memory.
White areas are yet to be calculated.  The lightly shaded (green) areas denote the current state
which must be held in memory, of order ${\cal O}{\rm log_2}N$ at time step $t=N$.  
\label{f:reblocking}
}
\end{figure*}

For this reason, we instead use the ``on-the-fly'' reblocking algorithm\cite{goddard} described below, which is efficient, robust, and lends itself easily to real-time signal processing during data collection.
A graphical representation of this method is shown in Fig.~\ref{f:reblocking}.
A block spin is defined to be the average spin at \emph{a particular site}, averaged over $m$ consecutive timesteps, where $m$ is the block size.
As the Monte Carlo simulation proceeds, we update the values of the most recent complete blocks of sizes that are powers of two: $m=1,2,4,8,16,\dotsc$.  We also accumulate the sums and sums-of-squares of block spins, so that we can ultimately construct the block averages and block variances.
We use an unbiased estimator for the variance, so that for, {\em e.g.},  a 16-element time series at block size $m=2$,
	\begin{align}
	\sigma_{(2)} &= 
		\tfrac{1}{8}
		\left[ \tfrac{\sigma_0+\sigma_1}{2} + \dotso + \tfrac{\sigma_{14}+\sigma_{15}}{2} \right]		
	\label{e:blockavg}
		, \\ 
	\left<(\Delta\sigma_{(2)})^2\right> &=
		\tfrac{1}{7}
		\left[ 
			\left( \tfrac{\sigma_0+\sigma_1}{2} \right)^2 
		+ \dotso 
		+ \left( \tfrac{\sigma_{14}+\sigma_{15}}{2} \right)^2
	 \right]		
		- 
		\tfrac{8}{7}
		\left[ {\sigma_{(2)}}  \right] ^2
.
	\label{e:blockvar}
	\end{align}
This procedure requires minimal computational effort,
and furthermore requires the storage of only 
${\cal O}({\rm log}_2 N)$ numbers in memory,
making the method amenable to real-time data analysis.
The variance $\left<(\Delta\sigma)^2\right>_m$
as a function of the \emph{temporal} block size $m$ can then be used 
to distinguish a telegraph signal that is free of spatial correlations
({\em i.e.}, one due to a stationary random telegraph signal)
from a telegraph signal due to an interacting model,
the {\em spatial} correlations of which 
contribute multiple timescales to each local switcher.

\begin{figure}[!htb]
\centering
\includegraphics[width=0.95\columnwidth]{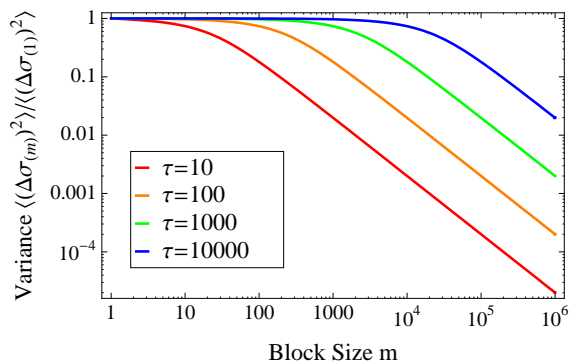}
\caption{
(Color online) Normalized block variances $\left<(\Delta \sigma_{(m)})^2\right>/\left<(\Delta \sigma_{(1)})^2\right>$ 
{\em vs.} block size $m$
of an uncorrelated switcher (stationary RTS) for various characteristic timescales $\tau$,
showing the crossover from constant to $1/m$ behavior.
\label{f:stationaryRTS}
}
\end{figure}
\begin{figure}[!htb]
\centering
\includegraphics[width=0.95\columnwidth]{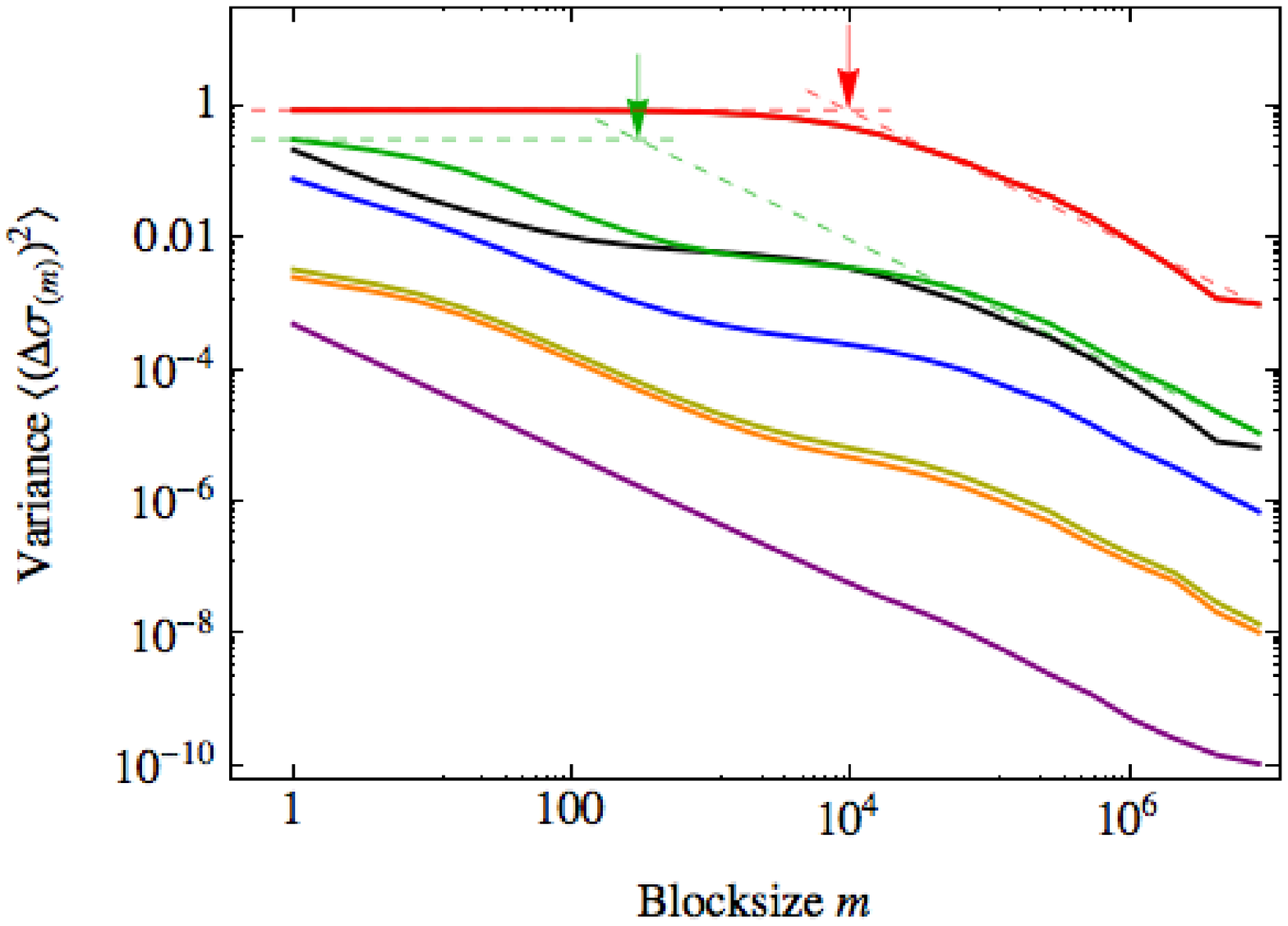}
\caption{
(Color online) 
Block variances $V_i^{(m)}$ {\em vs.} block size $m$ for 
the line of sites shown in Fig.~\ref{f:index2}.
The colors of the block variance curves correspond to the
colors of the arrows in Fig.~\ref{f:index2}.
For the red and green traces, the intersection of the dotted lines
shows the characteristic timescale of those sites, $\tau_i$, 
as described in the text.  
While the results for the red and purple traces mimic that of 
a stationary RTS with a long and short timescale, respectively, the other timetraces show
strong deviation from this behavior, 
a clear signature of the correlations in the system.
There are no brown and grey traces because the corresponding spins have zero variance
over the sampled time interval.
\label{f:blockvar}
}
\end{figure}

We first describe the results of this analysis for an
uncorrelated, independent switcher (a stationary random 
telegraph signal (RTS)), and then contrast it with local telegraph
noise from an interacting model, which encodes information
about the correlations.  At the block size $m$, the block variance
can be expressed as
\begin{eqnarray}
\left<(\Delta\sigma_{(m)})^2\right> &=& \frac{1}{m^2}\left< \sigma_{(m)}^2 \right> - \frac{1}{m^2}\left<\sigma_{(m)} \right>^2 \nonumber \\
&=& \frac{1}{m}\sum_{t=-m}^m (m-|t|)\psi(t)
\end{eqnarray}
where the average $\left<\cdots\right>$ is taken over all blocks of size $m$,
and $\psi(t)$ is the time-dependent part of the autocorrelation function $\Psi(t)$.
For a stationary discrete-time RTS\cite{machlup}, the autocorrelation is exponential (corresponding to a Lorentzian power spectrum):
	\begin{align}
	\psi(t) &= a e^{-|t|/\tau}
	\\
	\text{where}\quad
	a &= \frac{4pq}{(p+q)^2},
	\quad
	e^{-1/\tau} = \lambda = 1-p-q,
	\end{align}
where $p$ ($q$) is the conditional probability of switching from the ``up'' (``down'') state to the ``down'' (``up'') state at the next time step, and $\tau$ is the resulting characteristic timescale.
Thus the block variance is
\begin{equation}
\left<(\Delta\sigma_{(m)})^2\right> = a{m - m \lambda^2 + 2 \lambda(\lambda^m-1) \over m^2(\lambda-1)^2}
.
\end{equation}
Figure~\ref{f:stationaryRTS} shows the normalized block variance
$\left<(\Delta \sigma_{(m)})^2\right>/\left<(\Delta \sigma_{(1)})^2\right>$ {\em vs.} $m$
for uncorrelated switchers with various $\tau$.
The small $m$ behavior is a horizontal line on a log-log plot,
whereas the large $m$ behavior is a straight line corresponding to
$\left<(\Delta \sigma_{(m)})^2\right> \propto 1/m$.
The extrapolated intersection of these two linear regimes
may be used to extract a rough timescale $\tau_i$ for each site $i$,
	\begin{align}
	\tau_i &= m_\text{max} \left<(\Delta \sigma^i_{(m)})^2\right>/\left<(\Delta \sigma^i_{(1)})^2\right>
	\label{e:tau}
	\end{align}
where $m_\text{max}$ is the largest block size for which variance data is available.
We will see that deviations from the horizontal line
at low values of $m$ can be used as a local measure of 
spatial correlations.

In Fig.~\ref{f:blockvar}, we show the variance 
in our model, where spatial correlations have caused 
a deviation from the simple behavior of a stationary RTS.
The colors of the curves correspond to the colors
of the arrows in Fig.~\ref{f:index2}.
A couple of these curves mimic that of a stationary RTS.
For example, the red curve begins with the horizontal line characteristic
of a stationary RTS, 
and then crosses over to the $1/m$ behavior at a rather large timescale.
In addition, the purple curve displays the $1/m$ behavior throughout the plot,
consistent with a local stationary RTS with a rather short timescale. 
The rest of the curves in Fig.~\ref{f:blockvar}
show a striking deviation from the stationary RTS behavior,
in that they begin at small $m$ with a slope which is 
in between a horizontal line and $1/m$ behavior.
This may be taken as an indication of correlations.
In several of the curves, multiple ``knees'' are evident
within the same curve, further evidence of correlations,
in that there are multiple timescales present in the behavior of
a single site.   This type of analysis is amenable to 
real-time data processing of a telegraph STM signal in an ``active'' region,
in order to determine whether the noise is due to a 
local independent switcher (which may be due to, {\em e.g.}, a 
``dirt'' effect), or whether it is due to the more interesting case
of correlations due to interactions among the electronic degrees
of freedom.\footnote{Note that we have not focused on critical behavior
or on scaling behavior.\cite{dsfisher-prl,bray-moore-85}
Comparing those type of predictions to STM requires
that the measurements be averaged over a large field of view.
One advantage of our proposal is the focus on local physics,
without the need for spatial averages.}
In addition, it is free from the windowing errors that an FFT would entail.

In the right column of Fig.~\ref{f:kitchen-sink}, we plot maps of the local switching timescale,
$\tau$, estimated from Eqn.~\ref{e:tau}.
These may be viewed as a map of the dynamic clusters,
{\em i.e.}, those sites which fluctuate together.   
Note that this is different from the identification of the
low temperature domains.  
The behavior of the largest timescales as temperature is lowered
({\em i.e.}, in the active regions) is consistent with an Arrhenius law.
However, a single timescale is insufficient to
characterize the local dynamics because of the correlations,
and there are longer timescales in the active regions than
can be properly represented with a 2D color plot. 

\myhrulefill\section{Discussion}

We have argued that there should be telegraph noise in
STM experiments on locally striped materials
at intermediate temperatures.
Our proposal can be summarized as follows:  Using STM
in, {\em e.g.}, a tunneling asymmetry (TA) mode, 
the low temperature spatial map can be used 
to identify the low temperature domain structure.  
Then, place the tip on a domain boundary,
preferably where several small domains are clustered,
and take time traces of the signal at 
a few sites.  If no telegraph noise is evident, raise the temperature,
and repeat.  At high enough temperature, switching of the
stripe orientation should become thermally excited, 
resulting in local telegraph noise.  We have furthermore
proposed a criterion by which such noise can be analyzed
in the time domain, in order to determine whether it
is arising from a local, independent switcher, or whether it is
instead arising from correlated fluctuations in an interacting model.

Important issues include which signal should
be tracked, at what timescale do we expect telegraph noise to be evident,
and what are good candidate materials for testing our proposal. 
Some evidence of stripe structure has been reported in 
STM on BSCCO\cite{howald-pnas,fang} in the 8-15meV range, 
derived from measurements of dI/dV.
However, spatial modulations in the low energy response have
also been attributed  to quasiparticle interference.\cite{hanaguri,hoffman}
More dramatic evidence of locally unidirectional electronic domains,
with minimal signal processing involved,  
has been reported in Dy-BSCCO and NCCOC\cite{seamus-glass}  
via  tunneling asymmetry (TA) maps.  Specifically, the
ratio of the tunneling current at $\pm$150mV resulted in high contrast
images of domains with local Ising symmetry, consistent
with the model studied here.  For the purposes of 
studying noise associated with these Ising domains,
a time trace of the current near $\pm$150mV
would be a good place to start.
(Note that the magnitude of variation in signal in those experiments is
on the order of 30\%.) 
The variance test proposed here, which depends on
the time domain properties of a discrete signal change,
would apply whether the ``tip height'' $z$ was held constant
and a time trace of the current $I$ measured, or whether the
current $I$ was held fixed and a time trace of $z$ was recorded. 

It is worth emphasizing that telegraph noise has already been
reported in the transport properties of a YBCO nanowire.\cite{bonetti-04}  
In addition, similar behavior
was noted in a BSCCO nanowire\cite{dale-private},
although sufficient data was not gathered to demonstrate
telegraph noise in a time trace.
In this case, switches in the resistivity were observed
starting above $T_c$ (since the resistivity is zero below $T_c$),
and going as high as $T=150K$.  The noise was smaller
in magnitude, and less frequent than the corresponding noise
in YBCO.  However, BSCCO is much more amenable to 
an STM experiment, and given that locally unidirectional
domains have been observed in TA maps in Dy-BSCCO,
this system is a good candidate for observing
thermally excited stripe switching noise via STM.

In this paper we argue that fluctuations of stripe orientations will produce correlated telegraph noise in scanning tunneling experiments, and we propose a method for analyzing
such noise in the time domain.
This does not, of course, rule out the possibility of correlated noise emerging from other sources (e.g., side-to-side stripe fluctuations, or fluctuations of other inhomogeneous patterns, such as checkerboards or quasi-particle interference patterns).

\myhrulefill\section{Conclusions}
In conclusion, we have analyzed the local noise characteristics
of the random field Ising model in the disordered phase.
For materials with local stripe correlations 
in the presence of disorder, this model predicts that 
there will be thermally excited
fluctuations in the stripe orientation.  We propose that
such fluctuations should result in correlated telegraph noise in
the STM response near stripe domain boundaries as
temperature is raised.  
The use of this method in conjunction with TA scans
can establish whether there is a connection between
the observed low-$T$ unidirectional, glassy domains in Na-CCOC
and Dy-Bi2212, and the transport noise observed in
YBCO and LSCO, and also put constraints on our model. 
We have furthermore proposed that the block variance 
of telegraph noise can be used as a purely local
indicator of spatial correlations.
The criterion can be applied to any local probe
(AFM, MFM, {\rm etc.})  and because it is accumulated directly in the time domain,
it is amenable to real-time signal processing.

\myhrulefill\section{Acknowledgments}
It is a pleasure to thank G.~Aeppli, L.~Biedermann, 
J.~C.~Davis, E. Fradkin, T.~Hanaguri, E. Hudson, S.Kivelson, B.~Phillabaum, and 
M. Weissman for helpful discussions.
This work was supported by Research Corporation 
and by NSF Grant Nos. DMR 08-04748. and DMR 03-25939 ITR (MCC).


\end{document}